\documentclass[twocolumn]{aastex63}

\newcommand{\kms}{km s$^{-1}$ }

\received{XXX X, 2020}
\revised{XXX X, 2020}

\shorttitle{Simultaneous star and planet formation}
\shortauthors{Alves et al.}

\begin{document}

\title{A case of simultaneous star and planet formation} 
\correspondingauthor{Felipe O. Alves}
\email{falves@mpe.mpg.de}

\author[0000-0002-7945-064X]{Felipe O. Alves}
\affiliation{Max-Planck-Institut für extraterrestrische Physik,
Gie{\ss}enbachstra{\ss}e 1, 
Garching, 85748, Germany}

\author[0000-0003-2076-8001]{L. Ilsedore Cleeves}
\affiliation{Department of Astronomy, 
University of Virginia, 
Charlottesville, VA 22904, USA}

\author[0000-0002-3829-5591]{Josep M. Girart}
\affiliation{Institut de Ci\`encies de l'Espai (ICE), CSIC,
Can Magrans s/n, Cerdanyola del Vall\`es, 08193 Catalonia, Spain}
\affiliation{Institut d'Estudis Espacials de Catalunya (IEEC), 08034 Barcelona, Catalonia, Spain}

\author[0000-0003-3616-6822]{Zhaohuan Zhu}
\affiliation{Department of Physics and Astronomy,
University of Nevada, 
4505 South Maryland Parkway, Las Vegas, NV 89154, USA}

\author[0000-0003-2020-2649]{Gabriel A. P. Franco}
\affiliation{Departamento de F\'isica--ICEx--UFMG,
Caixa Postal 702,
30.123-970 Belo Horizonte, Brazil}

\author[0000-0002-5903-8316]{Alice Zurlo}
\affiliation{N\'ucleo de Astronom\'ia,
Facultad de Ingeniear\'ia y Ciencias, Universidad Diego Portales,
Av. Ejercito 441, Santiago, Chile}
\affiliation{Escuela de Ingenier\'ia Industrial,
Facultad de Ingenier\'ia y Ciencias, Unversidad Diego Portales,
Av. Ejercito 441, Santiago, Chile}
\affiliation{Aix-Marseille Universit\'e,
CNRS, LAM -- Laboratoire d'Astrophysique de Marseille,
UMR 7326, 13388 Marseille, France}

\author[0000-0003-1481-7911]{Paola Caselli} 
\affiliation{Max-Planck-Institut für extraterrestrische Physik,
Gie{\ss}enbachstra{\ss}e 1,
Garching, 85748, Germany}

\begin{abstract}
While it is widely accepted that planets are formed in protoplanetary disks, there is still much debate on when 
this process happens. In a few cases protoplanets have been directly imaged, but for the vast majority of 
systems, disk gaps and cavities -- seen especially in dust continuum observations -- have been the strongest 
evidence of recent or on-going planet formation. We present ALMA observations of a nearly edge-on ($i = 
75^{\circ}$) disk containing a giant gap seen in dust but not in $^{12}$CO gas. Inside the gap, the molecular 
gas has a warm (100 K) component coinciding in position with a tentative free-free emission excess observed 
with the VLA. Using 1D hydrodynamic models, we find the structure of the gap is consistent with being carved 
by a planet with 4-70 $M_{\rm Jup}$. The coincidence of free-free emission inside the planet-carved gap 
points to the planet being very young and/or still accreting. In addition, the $^{12}$CO observations reveal low-
velocity large scale filaments aligned with the disk major axis and velocity coherent with the disk gas that we 
interpret as ongoing gas infall from the local ISM. This system appears to be an interesting case where both a 
star (from the environment and the disk) and a planet (from the disk) are growing in tandem.  

\end{abstract}


\section{Introduction} 
\label{sec:intro}

One of the compelling aspects for our understanding of planet formation is the link between the synthesis of 
planetesimals, the growth of pebbles, and the emergence of gaps \citep{Chiang10,Johansen17}, where the 
latter seems to be a common feature in current observations of protoplanetary disks \citep{Andrews11, 
Andrews18,Francis20}. While planets are not the only explanation for gaps, they seem to be hard to avoid, 
especially for highly structured disks and particularly wide gaps. Models show that protoplanets can open wide 
gaps, tens of astronomical units across, with depleted dust and gas inside of them \citep[e.g.,][]{Zhu11}. Gaps 
are also now being observed in younger, Class I sources, and thus there is reinvigorated debate about {\em 
when} the process of planet formation starts \citep{Long17,Sheehan18,SCox20}. 

In the present paper, we report Atacama Large Millimeter/submillimeter Array (ALMA) continuum and 
molecular line observations of a unique system, [BHB2007]~1, a K7 young stellar object (YSO) with a 
bolometric luminosity of $\sim1.7$~L$_{\odot}$, effective temperature of 4060~K and a flat spectral energy 
distribution from near-infrared to mid-infrared bands (spectral class I $\rightarrow$ II) 
\citep{Brooke07,Forbrich09,Covey10}\footnote{The age estimation reported by \citet{Covey10} is based on a 
pre-{\it Gaia} distance of 130 pc. In addition, the evolutionary models used by these authors lead to two widely 
separated ages. The estimated stellar luminosity of this source is likely underestimated in previous works since 
our data show for the first time a highly inclined disk (\S \ref{almadust}).}. This indicates a source not older 
than 1 Myr, with most of the surrounding envelope dissipated and thermal emission arising from the 
circumstellar disk.  [BHB2007]~1 (BHB1, from now on) is located in Barnard 59 \citep[B59, distance $= 163 
\pm 5$ pc,][]{Dzib18}\footnote{The bolometric luminosity reported by \citet{Covey10} and mentioned above is 
scaled with the distance estimation reported by \citet{Dzib18}.}, the only site of active star-formation in the 
overall quiescent Pipe nebula. The core harbors a small cluster of YSOs \citep{Brooke07}, with BHB1 in a 
region with low interstellar extinction ($A_V\sim4$ magnitudes), about 5' ($\sim 0.24$~pc) west of the B59 
core \citep{Roman12}. 

As described in this paper, the system presents a complex morphology, with a clean and wide gap in the dust 
millimeter continuum, surprising for such a young object. Within the gap, there appears to be gas, and some 
kind of localized warm emission, seen also with the Karl G. Jansky Very Large Array (VLA). Furthermore, this 
system does not appear to be ``finished'' with accretion from the molecular cloud environment, as we see large 
scale, velocity-coherent filaments in the ALMA $^{12}$CO data. This paper presents the ALMA and VLA 
observations of this intriguing source, along with models to speculate on the nature of its hidden companion.  

\section{Observations}
\label{sec:obs}

\subsection{ALMA data}

BHB1 was observed with ALMA at 226 GHz as part of project code 2013.1.00291.S (PI: Alves).
The total continuum-dedicated bandwidth is 2.4 GHz. CASA~4.5 was used for calibration and imaging, where 
details of the calibration can be found in \citet{Alves17}. These observations used 44 antennae with baseline 
ranging from 15 to 1460 m. In this configuration, the maximum recoverable angular scale is $\sim$ 21\arcsec. 
We performed self-calibration of the {\it uv} visibilities by interpolating over decreasing solution intervals the 
phase gains (scan and integration time intervals) and amplitude gains (scan intervals). The reprocessed image 
has an {\it rms} noise of 0.095 mJy beam$^{-1}$, intensity peak of 19.5 mJy beam$^{-1}$ and flux density of 
0.16 $\pm$ 0.02 Jy. The continuum map was produced using Briggs robust parameter 0.5 and has a 
synthesized beam of $0.24\arcsec \times 0.20\arcsec$ and position angle (PA) of 77\degr (East of North).

$^{12}$CO and C$^{18}$O $J=2\rightarrow1$ were simultaneously observed covering the source and its 
surroundings. The spectral line data have velocity resolution of 0.35 km s$^{-1}$ ($\sim 270$ kHz) and peaks 
at channel $-0.95$ \kms, where intensity reaches 200 mJy beam$^{-1}$. Images were produced using a 0.5 
Briggs robust parameter. The final map has an rms noise of 3.5 mJy beam$^{-1}$ and a synthesized beam of 
$0.25\arcsec \times0.20\arcsec$ (PA $\sim 72\degr$). Both continuum and spectral line maps were primary 
beam corrected, important since the source is offset by $\sim 4.6\arcsec$ from the phase center.

\subsection{VLA data}
We have used the VLA in its most extended configuration to observe the 22.2~GHz continuum (K-band, $
\lambda = 1.35$ cm). A detailed description of the correlator setup is found in \citet{Alves19}, who reports on a 
distinct source with the same set of VLA observations used for BHB1. The VLA image was obtained using 
CASA clean and a robust weighting of 1, yielding a synthesized beam of $0.197''\times0.087''$ ($\simeq 21$ 
au)  with a PA of 10.3$^\circ$ and an {\it rms} noise of 7.5~$\mu$Jy~beam$^{-1}$.

\begin{figure*}[ht!]
\begin{center}
\includegraphics[width=1\textwidth]{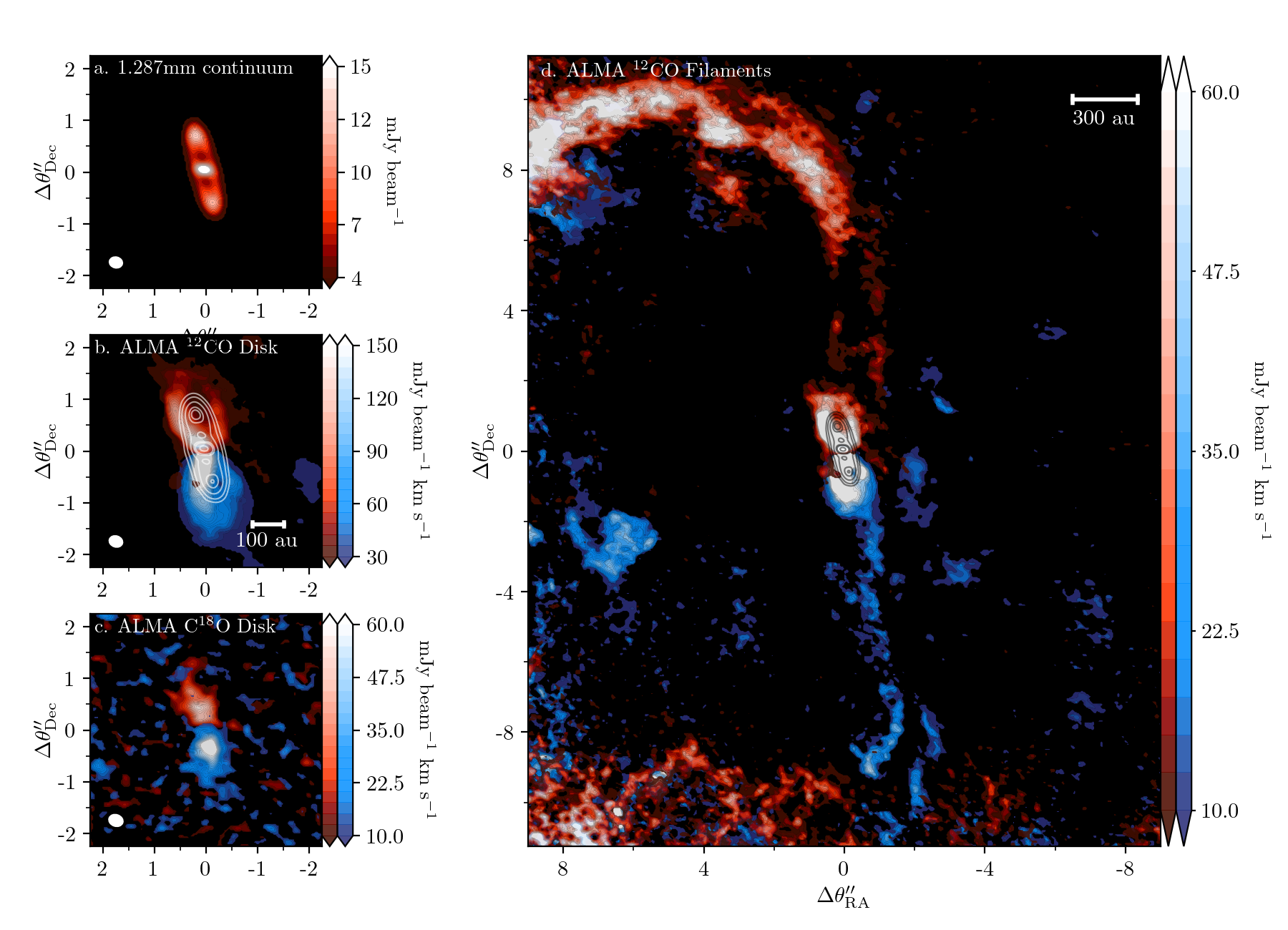}
\caption{{\it a}. ALMA continuum emission revealing the clear disk ring structure. Dust continuum contours are  
15, 30, 60, 100, 120, 180 times $\sigma$, the rms noise of the continuum. {\it Note:} the ``tilted'' asymmetry of 
the inner disk is likely due to the elongated beam. {\it b}. ALMA $^{12}$CO $J=2-1$ and {\it c}. C$^{18}$O 
$J=2-1$ disk emission. {\it d}. Large scale filamentary emission surrounding the [BHB2007] 1 disk. Red-shifted 
and blue-shifted emission is velocity-integrated between $4.5-7.0$ and $-0.4-2.7$~km~s$^{-1}$, respectively. 
Increased noise at the edge of the primary beam is more visible at the bottom-left hand corner, but the 
filaments in the image center and top are clear. 
\label{mainfig}}
\end{center}
\end{figure*}

\section{Results}

\subsection{Dust substructure in the disk}
\label{almadust}

The disk appears clearly in the millimeter data, with a quasi-symmetric morphology and three distinctive peaks, 
interpreted as an edge-on view of a gapped disk (Fig. \ref{mainfig}a). The disk has a radius of 107 au and is 
oriented $\sim 15\degr$ East-of-North. The inner and outer disk are separated by a gap with $\sim70$ au 
width. The mm dust brightness temperature peaks at $\sim10$~K.

\subsection{Disk CO observations} \label{sec:gas}

The disk's molecular gas traced by $^{12}$CO is more extended than the mm emission (Fig.~\ref{mainfig}b 
and d). The spatial distribution of the $^{12}$CO and C$^{18}$O velocity components show a clear Keplerian 
rotation pattern and a flared morphology. The $^{12}$CO has cloud contamination in the central channels; 
however, the C$^{18}$O (Fig.~\ref{mainfig}c) has less contamination and we find that the source velocity is 
consistent with previous estimates of 3.6~km~s$^{-1}$ \citep{Onishi99}. Using the uncontaminated channels 
and following the prescription of \citet{SSW16}, we fit a Keplerian model to the position-velocity diagram of the 
CO emission and find a mass of 2.23 $\pm$ 0.04 M$_{\odot}$ for the embedded protostar. 

\subsection{Large-scale filamentary CO emission }

Molecular emission reveals large scale ($\sim4000$ au) and narrow ($\sim80-300$ au) bipolar filaments 
connecting the ambient gas and the source (Fig.~\ref{mainfig}d).  These structures are distributed in a 
north-south orientation, similar to the major axis of the disk. The northern (red-shifted) and southern 
(blue-shifted) filaments exhibit a velocity shift of $\sim\pm2-3$~km~s$^{-1}$ with respect to the source ambient 
velocity.

No clear velocity gradients are seen along the filaments, whose entire structures are visible over a 
$\sim1$~km~s$^{-1}$ velocity range in each lobe.  This low velocity width implies we are likely seeing the 
filament along the plane of the sky. Interestingly, the pair of filamentary structures have global red and blue 
shifts. The northern filament is red shifted as the disk is with a similar velocity. The same is true for the 
southern filament, which is largely blue shifted, consistent with the disk rotation on the southern side of the 
disk.

If these features are indeed filaments, they appear to be orbiting the system like large propellers or gas 
streamers accreting into the disk (\S \ref{subsec:fil}). Alternatively, we could be seeing a limb-brightened large 
($>1000$ au) scale outer disk or flattened envelope, where the central channels are too optically thin to 
observe in emission. Additional observations of higher critical density tracers with deeper observations would 
be needed to disentangle a filament or remnant flattened envelope. What we can say with certainty is that we 
see large scale emission associated with the disk based on its velocities that is moving too slowly to be an 
outflow, and is oriented parallel to the disk plane rather than along a conventional outflow axis.

\begin{figure}[t]
\includegraphics[width=\columnwidth]{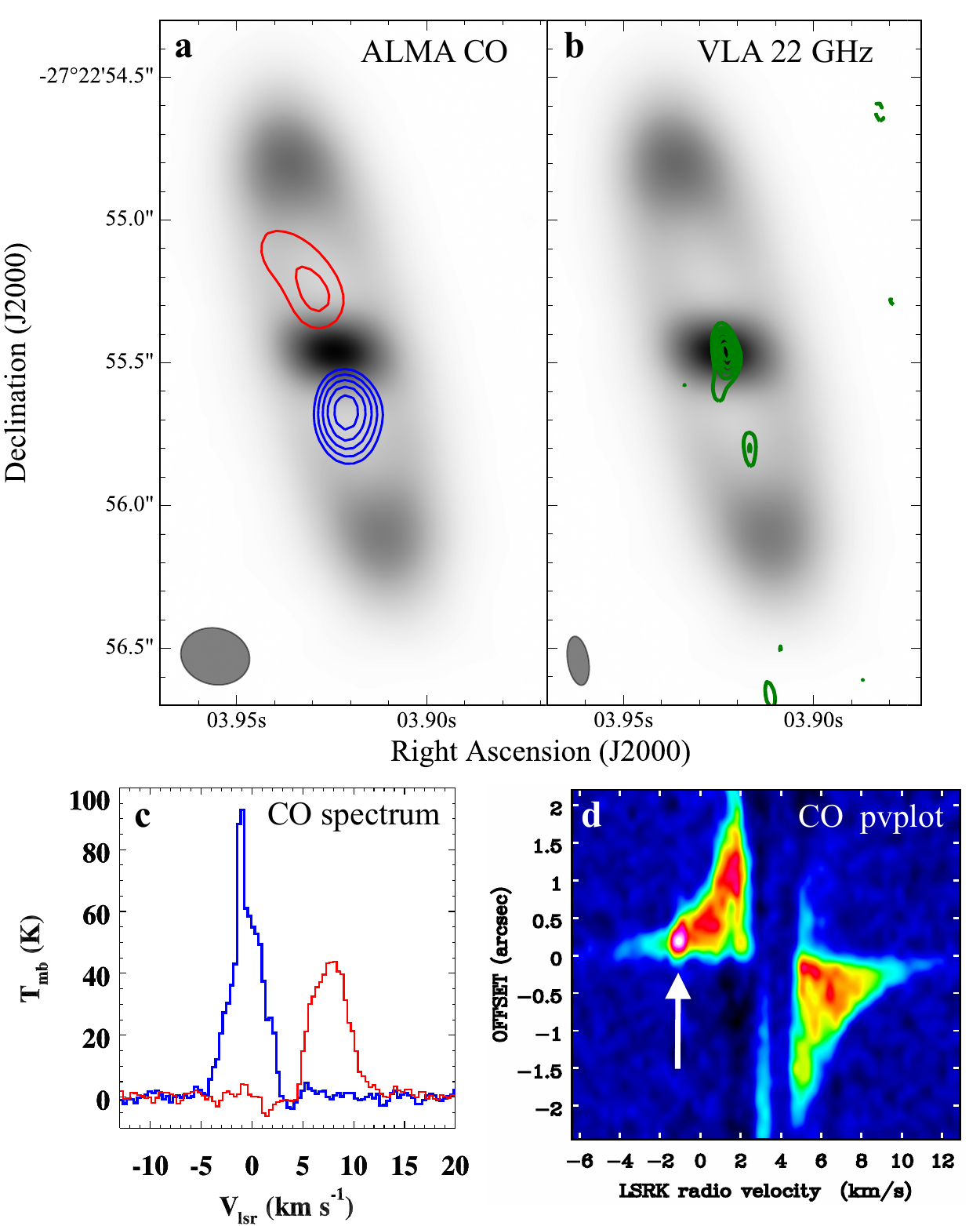}
\caption{{\it a.} ALMA dust continuum (grey scale) with CO emission (red and blue contours) overlaid. The 
contour levels are 25, 30 (red contours), 45, 50, 55, 60, 65, 70 (blue contours) times 3.5 mJy beam$^{-1}$, the 
$rms$ of the map, at a velocity offset of $\pm 4.6$ km s$^{-1}$ with respect to the systemic velocity. {\it b.} 
ALMA dust continuum with VLA emission (green contours) overlaid. The contour levels are $\pm$3, 4, 6 and 8 
times the $rms$ noise of the map (7.5 $\mu$Jy beam$^{-1}$). The synthesized beam of each contour map is 
shown in the lower left corner. {\it c.}  CO spectra taken from beam-size boxes centered on the dust minima in 
the north (red spectrum) and southern (blue spectrum) cavity. {\it d.} Position-velocity diagram (pvplot) taken 
from a cut along the disk major axis with a width of $\sim 2$\arcsec. The warm CO spot is indicated with an 
arrow.  
\label{figplanet}}
\end{figure}

\subsection{Localized emission in the south-east gap}
\label{subsec:local}

The VLA data show a compact radio source at the protostellar position with a flux density of 
$103\pm12$~$\mu$Jy (we refer to this source as VLA 1). In addition, the data reveal a marginal second peak 
in the southern gap at a separation of $0\farcs3$ ($\sim49$~au) from the star. This tentative source has a flux 
density of $30\pm8$~$\mu$Jy (for simplicity we refer to it as VLA 1b). The northern and southern gaps seen in 
the ALMA continuum map show intensity minima (5.21 and 5.28~mJy~beam$^{-1}$) that within uncertainties 
(0.10~mJy~beam$^{-1}$) are the same , i.\,e., there are no traces of VLA 1b at 1.3~mm.  This puts an upper 
limit for the spectral index of VLA 1b of $\la 0.8$ (assuming a 2-$\sigma$ upper limit at 1.3~mm), indicating 
that VLA 1b traces ionized gas.

Interestingly, VLA 1b coincides with a bright ($\sim 100$ K) and compact $^{12}$CO emission hot spot (Fig. 
\ref{figplanet}a,b). This spot is probably localized in the southern gap and VLA 1b since (1) it is spatially 
compact and with a narrow velocity width - both features indicate that it is truly compact in 3 dimensions - and 
(2) the peak velocity of this feature is close to the highest velocity at the projected radius (see 
Fig.~\ref{figplanet}c,d), i.\,e. the spot arises at a radius similar to the projected radius on the plane-of-the-sky.

\section{Analysis}

\subsection{Radiative transfer analysis}\label{sec:gapmeasure}

To estimate the size of the gap seen at mm-wavelengths, we explore parametric axisymmetric models using 
the TORUS radiative transfer code \citep{harriestorus2014}. Since the goal of this is to approximately constrain the size of the gap, we have adjusted the model parameters until we achieve a good fit by eye to the mm data and leave more complex modeling to a future analysis. 

The functional form of the disk model follows the \citet{lyndenbell1974} formalism for a viscously evolving disk, 
where
\begin{equation}\label{eq:sig}
\Sigma_{g} (R) = \Sigma_{c} \left( \frac{R}{R_c}  \right) ^{-\gamma} \exp{ \left[ -\left( \frac{R}{R_c} \right)^{2-\gamma} \right]}.
\end{equation} 
For the gas, $\Sigma_{c} = 15$ g cm$^{-2}$, $R_c = 100$~au, and $\gamma = 1$, and we do not include gas 
gaps. We assume a vertically Gaussian density distribution with a scale-height $h$ of 12~au at 100~au, and a 
flaring parameter of $\psi=1.2$ where $h=h_{100}\frac{R_{\rm au}}{100}^{\psi}$ and $h$ is in au. For the dust, 
we assume a global gas to dust ratio of 100, however the local gas/dust ratio in surface densities varies 
radially. The dust is split into two MRN population distributions with sizes up to 1~micron and 1~mm 
\citep[see][]{cleeves2016}, and both adopt pure astrosilicate compositions \citep{draine84}. The small dust 
grains follow the gas and contain 1\% of the dust mass. The mm-grains have the same functional form as 
Eq.~\ref{eq:sig}, but have a power law of $\gamma=0.2$ and an $R_c$ of 1000~au, well beyond the physical 
disk edge such that the mm dust surface density has a truncated power law behavior. The large grains' scale-
height is 10\% of the gas/small dust to approximate settling. The inner disk edge is set to 0.1~au and extends 
to 20 au. We model an empty gap, and then an outer disk component starting at 93~au extending out to 
160~au where it is truncated. We do not vary the depth of the gap as we only have one thermal continuum 
wavelength, and thus it is challenging to make accurate constraints on both the dust optical properties and the 
minimum dust mass in the gap, especially given its near edge-on nature. Instead we emphasize that this 
analysis is aimed to estimate the size of the gap, which we use to estimate possible companion masses for a 
range of uncertain physical properties (Section~\ref{hydromod}).

\begin{figure}
\includegraphics[width=.92\columnwidth]{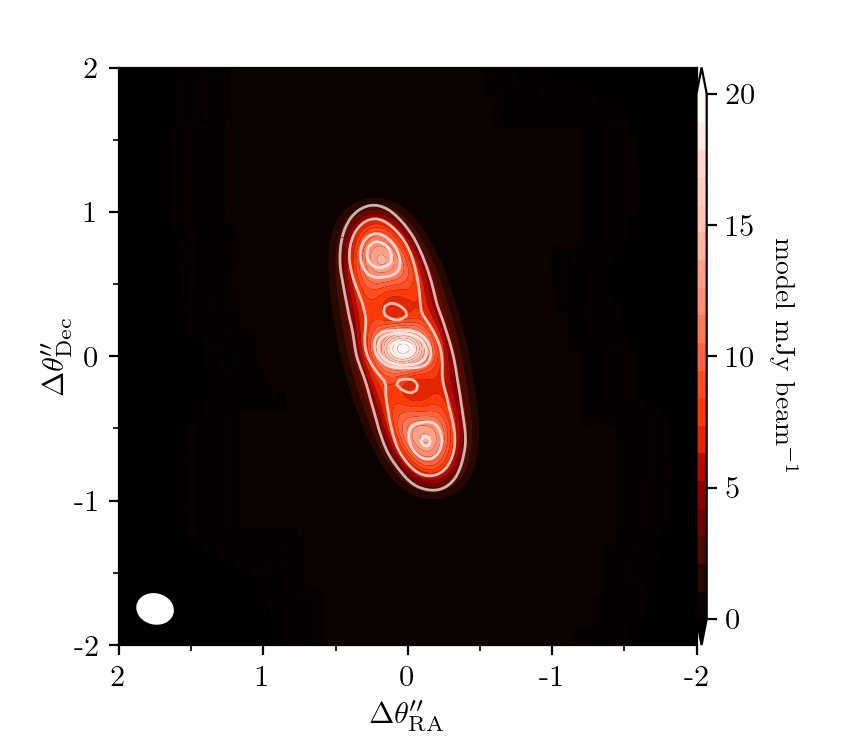}
\caption{1.287mm model continuum overlaid with the observed continuum in white contours drawn at the same 
significance levels as shown in Figure 1. The best model has a disk plane inclination of
$75\degr$ with respect to the plane of the sky.
\label{fig:dustmod}}
\end{figure}

Based on this structure and the stellar parameters described in Sections~\ref{sec:intro} and \ref{sec:gas}, 
TORUS computes a synthetic image at 1.287 mm, which we convolve with the observed beam. We explored 
just over 100 models and arrived at a reasonable fit as shown in  Figure~\ref{fig:dustmod}. To arrive at this fit, 
we varied all parameters except the $\gamma$ value of the gas, $R_c$ for the gas, the inner edge of the inner 
disk from 0.1 au, the relative dust mass between small and large grains, and the relative scale height of the 
large grains compared to the small. We find that within our grid of models, the inner edge of the outer ring can 
reasonably reproduce the structure qualitatively within $\pm5$~au. The gas+dust mass of this disk model is 
0.1~M$_\odot$. 

Taking this structure, we have additionally run a few non-LTE LIME \citep{Brinch10} models with our 
TORUS-derived structure to qualitatively investigate how much gas could be missing in the gap but not be 
clearly seen in the optically thick $^{12}$CO. Given the lower signal to noise of the C$^{18}$O observations, 
we do not try to compare simulations with these data. We assume a simple uniform CO abundance of 
$10^{-4}$ relative to H above dust temperatures of 20~K and $10^{-10}$ per H below similar to the approach 
of \citet{qi2006} to approximate CO freeze-out and photodissociation. We have artificially decreased the gas 
inside of the gap by varying factors. We find that a gap in the gas distribution is only visible when the gap 
depth is higher than a factor 100, otherwise it remains hidden due to the high optical depth of the $^{12}$CO 
emission. We emphasize that the true gap depth would require more detailed fitting, ideally with thermo-
chemical models, as the physical properties of gas in the gaps can be different depending on the gas to dust 
ratio in the gap, the presence or absence of small grains in the gap, the stellar UV heating of gas, and many 
more parameters 
\citep[e.g.,][]{bruderer12,bruderer13,bruderer14,vandermarel15,vandermarel16,vandermarel18,facchini18}, 
which are not constrained here. Instead, these simple models tell us that there is some amount of gas still in 
the gap and it is not fully cleared. Future deeper observations combined with thermo-chemical simulations 
should be conducted to better constrain the degree of gas depletion in the gap, if it exists.

\subsection{The Mass of the potential planet} 
\label{hydromod}

Such a wide gap in the dust may indicate the presence of a single or multiple planets. To estimate the mass of 
the potential planet, we adopt the approach used in DSHARP survey \citep{Zhang2018}. First, based on the 
gap profile, we estimate the relative width of the gap ($\Delta$ parameter in \citealt{Zhang2018}, which is the 
gap width over the radius of the outer gap edge) as 0.78. If we adopt DSD1 dust size distribution in DSHARP 
(dust size follows a power law distribution from 0.005 $\mu$m to 0.1 mm with a power law index of -3.5) and 
the gas surface density derived above, the Stokes number of the biggest dust particle at $50$ au is 
8.6$\times 10^{-4}$ and the derived K' parameter through the width-mass fitting  (defined in Table 1 of 
\citealt{Zhang2018}) is 0.40. Thus, with $(h/r)_{50 au}$=0.1, we can derive that the planet-to-star mass ratio is 
0.031 if the disk viscosity coefficient is $\alpha=10^{-3}$ and 0.015 if $\alpha=10^{-4}$. If we adopt DSD2 dust 
size distribution (dust size follows a power law distribution from 0.005 $\mu$m to 1 cm with a power law index 
of -2.5), the Stokes number at $50$ au is 0.086 and the derived K' parameter is 0.05. The planet-to-star mass 
ratio is 0.0039 if $\alpha=10^{-3}$, and 0.0019 if $\alpha=10^{-4}$. 

Taking into account all of these uncertain disk physical parameters, the possible planet-to-star mass ratio 
spans a wide range of [0.0019, 0.031]. Adopting the stellar mass computed in \S \ref{sec:gas}, the mass of the 
planet ranges between $\sim 4$ and 70 M$_{\mathrm{Jup}}$. We note that the dominant uncertainty in this 
determination is the maximum particle size and disk viscosity, rather than the details of the fit in 
Section~\ref{sec:gapmeasure}.  While the range is large, all masses point to a super-Jupiter sized companion, 
which would be consistent with the mass needed to launch jets capable of producing substantial free-free radio 
emission \citep{Zhu18}, see also \S \ref{subsec:local}.  .

\section{Discussion}

\subsection{Accretion filaments}
\label{subsec:fil}

The large scale filaments are aligned with the disk major axis and coherent with the Keplerian velocity of the 
protostellar disk seen in CO. In addition, no evidence of acceleration at along the filament makes it unlikely 
that the observed gas traces outflowing material. 

A more plausible explanation is that the filaments are streamers between the bulk of gas in the B59 core in the 
southeast and BHB1. The lack of a clear velocity gradient along the filaments suggests that the gas motion 
occurs mostly in the plane-of-sky.  Correspondingly, it is hard to assess which direction the gas flows; however, 
such low velocity gas is typical of infalling streamers observed toward accreting protostars \citep{Mottram13}.
In this context, the filaments could be molecular gas nurturing the disk and the YSO, which is consistent with 
single-dish observations of B59 showing dust and molecular patches protruding from the core toward the west, 
where BHB1 is located \citep{Duarte12}. 

\begin{figure}[t]
\includegraphics[width=.92\columnwidth]{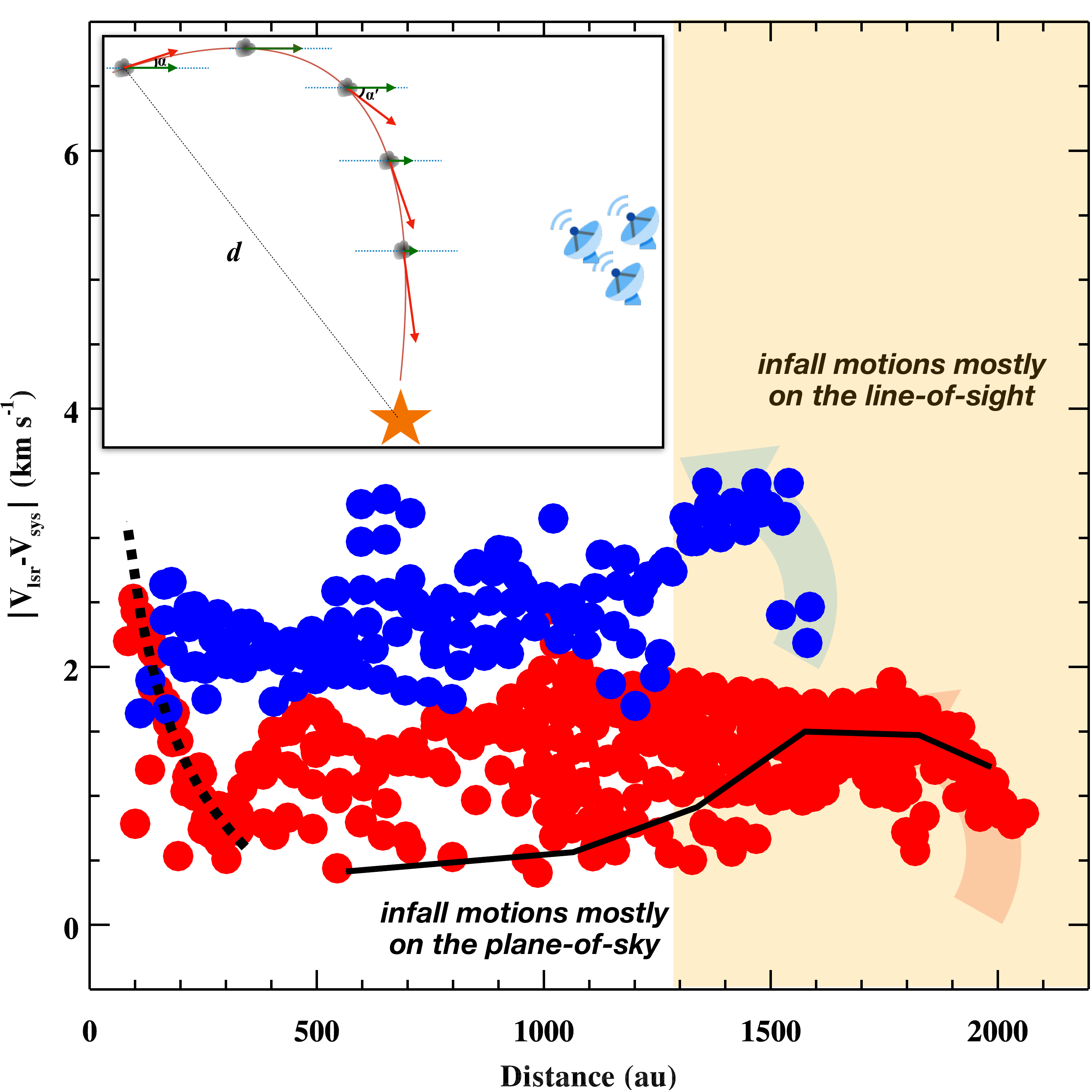}
\caption{Distance of CO ``pixels'' in the filament to the protostar. Red and blue points are pixels from the red- 
and blue-shifted components. These data were retrieved from a CO map smoothed to 1\arcsec\,resolution. 
The black line shows the infall gas model represented in the inset. The dashed line shows a Keplerian rotation 
model that matches the gas kinematics at distances shorter than 400 au, where the disk rotation dominates 
the kinematics.  
\label{figinfall}}
\end{figure}

Figure \ref{figinfall} displays the line-of-sight (LOS) velocity of CO gas as a function of distance to the protostar. 
At larger distances, the LOS velocity increases from $\sim 2000$ to 1750~au at the red-shifted lobe, and from 
$\sim 1650$ to 1500~au at the blue-shifted lobe. The LOS component then either decreases or becomes 
nearly constant at shorter distances. This velocity shift can be interpreted as material initially falling along the 
LOS, changing its direction and finally moving toward the source along the plane-of-sky. The region of 
increasing velocity is seen in Fig.~\ref{mainfig}d as bends in the endpoints of the filaments in both lobes. The 
bend is especially clear in the less filtered red-shifted lobe. Infalling gas parcels departing from rest with no  
energy dissipation have their LOS velocities determined as 
$V_{LOS} = \sqrt{2GM_{\star}/d}\times\cos{\alpha}$, where $d$ is the distance to the protostar, $\alpha$ is the 
angle between the velocity vector and the LOS and $M_{\star}$ is the protostellar mass (Fig.~\ref{figinfall}, 
inset). Large-scale accretion streamers have been recently reported by \citet{Pineda20}, who presented a 
scaled-up version of our filaments feeding a protostellar core. This indicates that the filamentary accretion 
streamers are the natural process in accreting objects regardless their age, as also indicated by time-
dependent multi-scale simulations \citep{Kuffmeier17}.  

\subsection{A giant planet hidden in the cavity?}

While the feature still must be confirmed, we find three intriguing lines of evidence that point to the possibility 
of a single ``dominant'' planetary or potentially brown dwarf companion in the disk. The first is the very broad 
disk gap apparent in the mm-emission. With a width of over 70~au centered around approximately 50 au, a 
massive gas giant or low mass brown dwarf would be necessary to clear such a gap. Based on the estimates 
presented in Section~\ref{hydromod}, its mass must be greater than 
4~M$_{\rm Jup}$ up to 70~M$_{\rm Jup}$. How such a massive object forms at this distance remains 
uncertain. Although there are cases of wide orbit giant exoplanets \citep[e.\,g.][]{Marois10}, the relative 
occurrence is not significant, and they are almost as rare as wide orbit brown Dwarfs \citep{Nielsen19}.

In addition to the disk morphology, there are two lines of evidence indicating preferential structure {\em inside} 
the southern gap. The first is  the presence of a CO ``hot spot'' at a position coinciding with the gap. Given that 
young planets are expected to be luminous and thereby heat their surroundings, this may be a sign of such 
localized heating \citep{wolf2005,Cleeves15}. The second line of evidence is the presence of marginal radio 
emission, VLA~1b, inside the southern gap. No such source is seen in the ALMA data, suggesting that the 
emission originates from ionized gas (see Section~\ref{subsec:local}). It is common to see radio emission 
associated with strong winds and/or jets from protostellar objects  \citep{Hull16, Anglada18}. There is a known 
correlation between radio luminosity and a YSO's bolometric luminosity. If VLA~1b is of the same kind, its radio 
flux would be consistent with a substellar object \citep{Morata15,Rodriguez17,Ricci17}. If confirmed, VLA~1b 
would be a signpost that there is a substellar object significantly accreting gas from the disk. Alternatively, the 
radio emission could be produced by the a strong magnetosphere of a fast rotating (sub)stellar object. 
However, the radio luminosities observed in old brown dwarfs are $1-2$ orders of magnitudes lower than the 
flux detected here \citep{Berger01,Kao18}. Yet, these non-thermal mechanism possibilities can be variable, 
and so repeat follow up observations would likely be necessary in ascertaining the nature of VLA~1b. Near-
infrared observations from scattered light produced by the stellar radiation are also being analysed in order to 
constrain the nature of the companion (Zurlo et al. in preparation).

\section{Conclusions}
We report the discovery of a disk with a wide gap, even though the disk itself still appears to be fed by 
extended filaments detected in molecular gas. As a result, this system asks the question, can planets form 
before the disk itself is fully formed? Furthermore, these data put new time constraints on the giant planet 
formation process, if indeed they form so early ($<1$~Myr). Our observations present a detailed view of a 
circumstellar disk, with bright thermal emission from the inner and the outer disk and a large zone of depleted 
dust between them. Locally, compact and warm gas is detected within the dust gap, coinciding in position with 
centimeter-wavelength radio emission. Our data are well represented by a model of a protoplanetary disk 
carved by a giant planet or brown dwarf from which bright non-thermal emission is produced. 

\acknowledgments
The authors thank the anonymous referee for the very insightful report. F.O.A. and P.C. acknowledge financial support from the Max Planck Society. L.I.C. acknowledges support from the David and Lucille Packard Foundation and NASA ATP 80NSSC20K0529. J.M.G. is supported
by the grant AYA2017-84390-C2-R (AEI/FEDER, UE). 
G.A.P.F. acknowledges support from CNPq and FAPEMIG (Brazil). A.Z. acknowledges support from the FONDECYT {\it Iniciaci\'on en investigaci\'on} project number 11190837.  This paper makes use of the following ALMA and VLA data: DS/JAO.ALMA\#2013.1.00291.S, VLA/16B-290.
ALMA is a partnership of ESO
(representing its member states), NSF (USA) and NINS (Japan), together with NRC (Canada), NSC and ASIAA
(Taiwan), and KASI (Republic of Korea), in cooperation with the Republic of Chile. The Joint ALMA Observatory is
operated by ESO, AUI/NRAO and NAOJ. 
\facilities{ALMA, VLA}

\bibliographystyle{aasjournal}
\bibliography{refalves}{}

\end{document}